\documentstyle[preprint,aps,pre]{revtex}
\begin{document}
\draft
\title{Coherent oscillations and incoherent tunneling in one - dimensional
asymmetric double - well potential}
\author{V.A.Benderskii $^{1 , 2}$ \, , E. I. Kats $^{3}$ \footnote{
Also, from  L. D. Landau Institute for Theoretical Physics, RAS,
117940, Kosygina 2, Moscow, Russia.}}
\address{$^1$ Institute of Problems of Chemical Physics,
RAS \\
142432 Moscow Region, Chernogolovka, Russia; \\ 
$^2$ Lab. Spectrometrie Physique, UJF \\
BP 87, St. Martin d'Heres, Cedex, France; \\
$^3$ Laue-Langevin Institute, F-38042, Grenoble, France.}
\date{\today}
\maketitle

\begin{abstract}
For a model 1d asymmetric double-well potential we calculated so-called
survival probability (i.e. the probability for a particle initially localized in one
well to remain there). We use a semiclassical (WKB) solution of the Schroedinger equation.
It is shown that behavior
essentially depends on transition probability, and
on one dimensionless parameter $\Lambda $ which is 
a ratio of characteristic frequencies for low energy
non-linear in-well oscillations and inter wells tunneling. 
For the potential describing a finite motion (double-well)
one has always a regular behavior.
For $\Lambda \ll 1$ there
is well defined resonance pairs of levels and the survival probability
has coherent oscillations related to
resonance splitting. However for $\Lambda \gg 1$ no oscillations at all
for the survival probability, and there is almost an exponential decay with
the characteristic time determined by Fermi golden rule.
In this case one may not restrict oneself to only resonance pair levels.
The number of perturbed by tunneling levels grows proportionally to $\sqrt \Lambda $
(by other words instead of isolated pairs there appear the resonance regions
containing the sets of strongly coupled levels).
In the region of intermediate values of $\Lambda $ one has a crossover between
both limiting cases, namely the exponential decay with subsequent long period
recurrent behavior.
\end{abstract}


\section{Introduction}

Double level systems and models appear in various contexts in physics, chemistry
and biology. 
The recurrent interest to the topic is related mainly with
fairly rich and interesting physics of the systems, and
with the experimental activity on several
classes of systems which can be viewed as good physical realization
of double level models (including fashionable quantum dots, see e.g. \cite{YF99}).
Among the possible types of behavior we will
particularly be concerned with coherent oscillations and
incoherent (dissipative like) tunneling.
Our goal is to propose a simple mathematical model to illustrate
crossover from coherent oscillations  to dissipative tunneling 
(decay or relaxation),
which are also related to incoherent transitions in multidimensional
oscillator systems.
In a certain sense this crossover reveals many features of chaotic behavior.
It is a common wisdom now that classical chaos is defined as extreme
complexity of the trajectories in phase space, with the trajectories being
very sensitive to small changes in the initial conditions \cite{ZC79}, \cite{LL83}.
As well evidently that the state vector (wave function) of a closed
quantum system strictly speaking 
does not exhibit chaotic motion, as a consequence of the unitary nature
of time evolution. But in fact since in quantum mechanics trajectories
in the phase space can not be introduced due to Heisenberg uncertainty
principle, the standard classical concept of the stability becomes ambiguous
(see e.g.\cite{WJ82}, \cite{BE1}, \cite{BE2},
\cite{NA}, \cite{KH00}).

We put forward a simple (but yet non-trivial)
model of 1d asymmetric double-well potential which can be used
to describe under relatively weak assumptions a crossover 
from coherent oscillations (say mechanical behavior) to incoherent decay or dissipative tunneling
(say ergodic behavior).
The essential part of the model we will present is to illustrate this
semiclassical quasi-chaotic behavior.
In fact the illustration was made long ago by Fermi, Pasta, and Ulam \cite{FPU}.
They performed computer studies of energy sharing and ergodicity for weakly
coupled systems of $N$ oscillators. Later on, the results of \cite{FPU} were
confirmed and refined (see e.g. \cite{JA63}, \cite{FW63}).
But all these papers were devoted to systems with many degrees of freedom
($N \gg 1$ dimensional phase space) for the cases where the motion is nearly
integrable and irregular in different energy regions.
Level statistics for such kind of mixed systems
(i.e. when behavior is regular and chaotic 
in different phase space regions) changes gradually from Poisson to Wigner
type of distributions \cite{BR84}, \cite{CG86}, \cite{ZM86}. 
Thus these systems become non-integrable when the energy exceeds a certain critical value.
Just on the contrary we will propose and investigate in 1d a conservative system
with time independent Hamiltonian which is evidently always integrable,
and it does not generate classical chaos. 

For the sake of completeness let us note that the tunneling in the mixed (i.e.
regular-chaotic) systems has been studied as well for two level systems 
when one of the levels interacts with a chaotic state \cite{TU94}, \cite{SW01} (see also
review \cite{BT93} and references therein). In the case of a resonance between
the tunneling doublet and suitable chaotic states, the tunneling is enhanced
(so-called chaos assisted tunneling) and has very strong resonance dependence
on quantum numbers. Similar effects due to transverse vibrations
take place for isolated Fermi resonances
in tunneling systems \cite{BV99}.

Our paper has the following structure. Section II contains basic equations necessary
for our investigation. Section III is devoted to the
calculation of so-called survival probability.
We use the semiclassical approach \cite{LL} (see also \cite{b1} and references
herein).
The last section IV contains the summary.
The appendix to our paper is devoted 
to the technical and methodical details of the calculations.

\section{Asymmetric 1d double-well potential}

The simple model studied in this paper consists of a quantum particle
in one dimensional asymmetric double-well potential $U(X)$ with one-parameter dependent shape.
Using the tunneling distance $a_0$ and the characteristic frequency of the
oscillations around the left minimum $\Omega _0$, we can introduced the
so-called semiclassical parameter  
$\gamma \equiv m\Omega _0 a_0^2/\hbar  \gg 1 $ ($m$ is a mass of a particle,
and further we will set $\hbar = 1$ measuring energies in the units of
frequency), 
which is assumed to be sufficiently large, i.e. the tunneling matrix
element should be small in $\Omega _0$ scale. 
The choice of the model potential is dictated by the principle
of minimal requirements. Our aim is to describe in
the frame work of one universal model crossover from 
symmetric double-well potential to so-called decay potential,
and to do it we need a parameter to make the right well ($R$-well) 
deeper and wider than the left well
($L$-well).

Using $\Omega _0$ and $a_0$ to set corresponding scales,
the model potential satisfying these minimal
requirements 
can be written in the following dimensionless form
\begin{eqnarray}
\label{b1}
V(x) = \frac{1}{2}x^2(1-x)\left [1 + \frac{1}{b^2}x\right ] \, ,
\end{eqnarray}
where $V \equiv U/(\Omega _0 \gamma )$, and $x \equiv X/a_0$.
The dimensionless parameter $b$ allows us 
to change the shape of the right well ($R$-well), and to consider both limiting cases, namely
a traditional symmetric double-well potential (for $b =1$), and 
for $b \to \infty $ a decay potential (or by other words to change
the level spacings from $\Omega _0^{-1}$ scale to zero ).
In fact it can be shown (see below and the Appendix to the paper) that qualitatively
all our results do not depend on the concrete form of
the one parametric potential satisfying these requirements (only on the density of $R$-states).
Behavior in both limiting cases are well known, and for $b=1$ one has
coherent quantum oscillations, typical for any two-level systems, 
while for $b \to \infty $ there is
a continuum spectrum of eigen states for $x \to +\infty $ and one can find 
an ergodic behavior (incoherent decay). Our main goal in this section
is to study crossover between both limits at variations of $b$.

The general procedure for searching semiclassical solutions of the Schroedinger
equation with the model potential (\ref{b1}) 
has a tricky point. The fact is that in the $L$-well we have a discrete
eigenvalue spectrum (stationary states) while for the $R$-well in the case $b \gg 1$
we have quasi-stationary states, which are characterized by wave
functions $\Psi _n(X)$ exponentially increased in the region of $\varepsilon \gg V(X)$.
Both kind of states are defined on different sheets of complex energetic surfaces \cite{LL},
and to treat both kind a states one should use different tools, namely, 
the standard quantization of the stationary states from the discrete part of the
spectra \cite{LL}, and proposed long ago by Zeldovich \cite{zeld}
for quasi-stationary states 
the flux probability conservation law, which leads to the Lorentzian envelope
for spectral distribution functions. Unlike \cite{zeld} in our case
we get the Lorentzian envelope filled by $\delta $-peaks of the final states.

The procedure is described in the Appendix, and it
includes three steps (see \cite{LL}, \cite{zeld},
and we will use notations from \cite{b1}):
\begin{itemize}
\item
First one should find the action $W_L$ in the classically allowed region
(i.e.$W_L$ between turning points) in the left
well ($L$-well), and apply the semiclassical quantization.
For the low energy states in the $L$-well it leads to the following relation
\begin{eqnarray}
\label{b2}
\gamma W_L = \pi \left [n + \frac{1}{2} + \chi _n \right ] \equiv \pi \varepsilon _n \, ,
\end{eqnarray}
where, integer numbers $n$ numerate eigenvalues,  $\chi _n$ is
determined
by an exponentially small 
phase shift, and the last r.h.s. of (\ref{b2}) is
in fact the definition for eigenvalues $\varepsilon _n$.

\item
Second, the same should be done for the right well ($R$-well).
The calculation is almost trivial in the limit $b \gg 1$ (when the potential (\ref{b1})
becomes strongly asymmetric)
\begin{eqnarray}
\label{b3}
\gamma W_R = \gamma W_R^{(0)} + \pi \beta \varepsilon
\, ,
\end{eqnarray}
where the dimensionless energy $\varepsilon $
is counted from the bottom of the $L$-well,
the action $W_R^{(0)}$ is
\begin{eqnarray}
\label{b4}
\gamma W_R^{(0)} = \frac{\pi }{16 b}(b^2 - 1)^2(b^2 + 1)
\, ,
\end{eqnarray}
and
\begin{eqnarray}
\label{b5}
\beta  = \frac{b^2 +1}{b} \simeq b \, , \, {\rm for}  \quad b \gg 1
\, ,
\end{eqnarray}
Note that the parameter $\beta = 2\Omega _0/\omega _R $ is
proportional to the density of states in the $R$-well ($\omega _R$ is the frequency 
of non-linear oscillations in $R$-well at $\varepsilon = 0$), and
therefore knowing the magnitude $\beta $ one can compute the density
of states in the $R$-well, which grows proportional to $b$ for $b \gg 1$.
It is convenient to rewrite (\ref{b3}) - (\ref{b4}) in the same form as
(\ref{b1})
\begin{eqnarray}
\label{b6}
\gamma W_R  = \pi \left [n_R + \frac{1}{2} + \alpha _n + \beta \chi \right ]
\, ,
\end{eqnarray}
where $n_R$ and $\alpha _n$ are integer and correspondingly fractional
parts of the quantity
\begin{eqnarray}
\label{b7}
\frac{\gamma W_R^{(0)}}{\pi} + \beta \left (n + \frac{1}{2} \right ) - \frac{1}{2}
\, .
\end{eqnarray}
The physical meaning of $\alpha _n$ is the deviation from a resonance
between the $n$-th level in the $L$-well and the nearest level
in the $R$-well. By the definition of a fractional part $|\alpha _n| < 1/2\beta $.

\item
And as the last step, again using the quantization rule, one can find the spectrum.
\end{itemize}

It turns out (see Appendix)
that the spectrum and the behavior of the system depends crucially on the
parameter $\Lambda \equiv \beta R_n$, where
\begin{eqnarray}
\label{b8}
R_n = \frac{2^{n+2} \gamma ^{n +1/2}}{\pi ^{1/2} n!} \exp (- 2 \gamma W_B)
\end{eqnarray}
is the $\beta $ independent decay rate of the $n$-th metastable state of the
$L$-well at $b \to \infty $ ($W_B$ is the action in the classically forbidden
(between turning points) region).

For $\Lambda \ll 1$ solving the quantization relation (\ref{A1}),
one can easily find
\begin{eqnarray}
\label{b9}
\varepsilon _{n \pm} = n + \frac{1}{2} \pm \frac{1}{2 \beta } \left [\sqrt {\alpha _n^2 + \frac{4}{\pi ^2}
\Lambda } - \alpha _n \right ] \, 
.
\end{eqnarray}
This expression (\ref{b9}) determines the resonance pairs of the levels,
so-called two-level systems.

Besides from the same quantization rule (\ref{A1}) we get analytically
(i.e. for arbitrary values of $\Lambda $) eigenvalues for the
$R$-well in the vicinity of the resonance doublet
\begin{eqnarray}
\label{b10}
\varepsilon _{n m} = n + \frac{1}{2} + \frac{1}{2 \beta } \left [\sqrt {(m - \alpha _n)^2 + \frac{4}{\pi ^2}
\Lambda } - (m - \alpha _n)\right ] \quad  m = \pm 1 , \pm 2 , .... 
\end{eqnarray}
These levels are numerated by the quantum number $m$.

For $\Lambda \ll 1$ all displacements of the levels due to tunneling
are small, and two-level system approximation is valid
(i.e. there is well defined isolated resonance pairs of levels with 
splitting $\propto (R_n
/\beta )^{1/2}$).
The situation becomes completely different for $\Lambda \geq 1$.
In the limit $\Lambda \gg 1$ we get almost equidistant spectrum of mixed $L - R$ levels in 
the vicinity
of the following values of $\chi $ (see Appendix for the details)
\begin{eqnarray}
\label{b11}
\chi \equiv \chi _{n m} = \pm \frac{m + 1/2 - \alpha _n}{\beta }\left 
[1 + \frac{1}{\pi  \Lambda }\right ]
\, .
\end{eqnarray}

The given above expressions (\ref{b10}) - (\ref{b11}) show that the number 
of the perturbed
by tunneling levels grows proportionally to $\sqrt \Lambda $. In Fig. 1 
we have shown
the displacements of the levels perturbed by tunneling. 
These displacements are decreased very rapidly for the levels with quantum
numbers larger than $\sqrt \Lambda $. The scales in this figure are given by the semiclassical
parameter $\gamma $ which relates to the $L$-well and the barrier. Once the scales are fixed
the $R$-well is characterized by the eigenfrequency $\propto 1/b$ at $\varepsilon = 0$ (or what
is the same by the density of states or by the action $W_R$ in the $R$-well).

Summarizing the results of this section, thus we have shown that instead of
isolated two level systems taking place for $\Lambda \ll 1$, in the opposite
limit $\Lambda \gg 1$ there appear the resonance regions containing the sets
of strongly coupled levels. The resonance widths are determined by tunneling
matrix elements ($H_{1 2}^2 = \omega _L \omega _R \exp (- 2\gamma
W_B)/4 \pi ^2
= R_n/\beta $). In
spite of the fact that 
for any finite values of $\Lambda $ (and $b$) we have only the discrete
spectrum of real eigenstates, found above mixing of $L - R$ states very
closely resembles the representation of quasi-stationary states in terms of eigenstates of a continuous
spectrum. This behavior can be formulated by other words in terms of the so-called
recurrence time, i.e. the characteristic time when the system is returned to the initial
state. For a finite motion (i.e. for a finite value of $b$) the behavior
of the system remains regular. The recurrence time (i.e. in the case merely
coherent oscillation period) is proportional
to $1/H_{1 2}$ for $\Lambda \ll 1$, while for $\Lambda \gg 1$ this time
scales as $1/\omega _R$ (as a long-period time scale).

\section{Survival probability}

The tunneling dynamics can be characterized by the time evolution
of the initially prepared localized state $\Psi (0)$, and by the definition
the survival probability of the state is
\begin{eqnarray}
\label{b12}
P(t) \equiv |\langle \Psi (0) | \Psi (t) \rangle |^2
\, .
\end{eqnarray}
For the stationary states evidently $P(t) = 1$, while for quasi-stationary
(decaying states), the survival probability reads
\begin{eqnarray}
\label{b13}
P(t) = \exp (- \Gamma t)
\, ,
\end{eqnarray}
where $\Gamma $ is the decay rate which should be found, and we use
$\omega _R^{-1}$ for the time scale.

The simplest case is the coherent tunneling dynamics of two-level states.
Let us consider the $ n - n^\prime $ resonance region.
The eigenfunctions of isolated $R$ and $L$ wells, $\Psi _n^L$, and $\Psi _{n^\prime }^R$.
If one has the initial state 
$$
\Psi (0) = \Psi _n^L  \, ,
$$
the survival probability can be easily calculated
\begin{eqnarray}
\label{b14}
P(t) = \frac{1}{2}\left [1 + cos\left (2t\sqrt {\frac{R_n}{\beta }}\right )\right ]
\, .
\end{eqnarray}

Normalized wave functions in the $L$-well can be calculated
trivially, and using standard semiclassical wave functions for the $R$-well,
we are in a position  to compute the survival
probability for a general case as a function of $\Lambda $.
The results are shown in Fig. 2.

For $\Lambda \ll 1$, $P(t)$ oscillates with characteristic time scales
proportional to $H_{12}^{-1} = \sqrt {\beta /R_n}$.
In the region $\Lambda \simeq 1$ these oscillations are strongly suppressed.
The reason for the suppression of oscillations is related to interference of the states
with energies in the resonance region. As a result of the interference the total
probability for the system to return back from the $R$-well is decreased, and low-frequency
modulation of coherent tunneling is raised. The period of the modulation
grows with $\beta $, and in the limit $\Lambda \gg 1$ we get the dense spectrum of
states in the $R$-well, and almost exponential decay for $P(t)$ with $\beta $-independent
relaxational time $\tau \propto R_n^{-1}$.
In this case the survival probability (i.e.
the probability to keep the system in its initial state) for the time
interval $ \ll 1/\omega _R$ decay almost exponentially with time, and
the characteristic relaxation time $\tau $ is determined by Fermi golden rule,
i.e. $\tau ^{-1} \propto H_{1 2}^2/\omega _R$.
This result is also conformed to van Hove statement \cite{HO55} concerning
quasi-chaotic behavior of semi-classical systems at time scales of the order
of $\omega _R/H_{1 2}^2$.

We can relate the phenomenom described above (i.e. almost
vanishing probability for back-flow from the $R$ to $L$ well) to
the Fermi golden rule for a transition probability
\begin{eqnarray}
\label{b15}
W_{fi} = 2\pi |H_{f i}|^2 \rho _f 
\, ,
\end{eqnarray}
where $H_{i f}$ is the matrix element between the initial state
$E_i$ and the final state $E_f$, and $\rho _f$ is the density of final states.
For our case ($H_{i f} \equiv H_{12} = \sqrt {R_n/\beta }$, and
$\rho _f = \beta /2$) we get easily
$$
W_{i f} = \pi R_n \, ,
$$
which does not depend on $\rho _f$.
Therefore the Fermi golden rule corresponds to the limit
when the back flow from the $R$-well is totally suppressed due to the interference.

The survival probability can be related also to spectral distribution
of the initially localized in the $L$-well states.
Indeed, by the definition of the spectral distribution $S(E)$
of the initially prepared localized state is determined by the transition
amplitudes in expansion over the eigenstates $(\Psi _n , E_n)$:
\begin{eqnarray}
\label{b16}
S(E) = \sum _{n}|\langle \Psi (0) | \Psi _n \rangle |^2 \delta (E - E_n)
\, ,
\end{eqnarray}
and therefore
\begin{eqnarray}
\label{b17}
\langle \Psi (0) | \Psi (t) \rangle = \int _{-\infty }^{+\infty } S(E) \exp (- i E t) dE
\, .
\end{eqnarray}
For $\Psi (0) \equiv \Psi _i^L$ the spectral distribution is a set of $\delta $-peaks with
Lorentzian envelope
\begin{eqnarray}
\label{b18}
S(E) = \frac{2}{\pi } \frac{\sqrt {R_i\beta }}{\beta (E - E_i)^2 + R_i}\delta (E - E_i)
\, .
\end{eqnarray}
Crossover from the coherent oscillations to exponential decay occurs when
the Lorentzian envelope begins to fill up by $\delta $-peaks of the final states.
Note that the width of the Lorentzian envelope (\ref{b18}) does not depend on the
final state density (see Appendix and also \cite{zeld}).
We have shown the results of the calculation of the spectral distribution in Fig. 3.

\section{Conclusion}
Let us sum up the results of our paper.
We investigated the behavior of a quantum particle in 1d asymmetric double-well
potential with one parameter dependent shape, which allows us to consider
in the frame work of one universal model the crossover from the traditional
symmetric double well potential to the decay one.
We have shown that behavior
essentially depends on transition probability, and
on dimensionless parameter $\Lambda $ which is 
a ratio of characteristic frequencies for low energy
non-linear in-well oscillations and inter wells tunneling. 
For the potential describing a finite motion (double-well)
strictly speaking one has always a regular behavior.
For $\Lambda \ll 1$ there
is well defined resonance pairs of levels and the survival probability
has coherent oscillations related to
resonance splitting. However for $\Lambda \gg 1$ there are 
no oscillations at all
for the survival probability, and there is almost an exponential decay with
the characteristic time determined by Fermi golden rule.
In this case one may not restrict oneself to only resonance pair levels.
The number of perturbed by tunneling levels grows proportionally to $\sqrt \Lambda $
(by other words instead of isolated pairs there appear the resonance regions
containing the sets of strongly coupled levels).
In the region of intermediate values of $\Lambda $ one has a crossover between
both limiting cases, namely the exponential decay with subsequent long period
recurrent behavior.

However a number of remarks related to our results are in order.
Many features often classified as evidences of quantum chaos in fact as we have
illustrated in our model can occur for well defined states possessing only 
discrete energy levels. The deviation from two level system behavior,
taking place for $\Lambda \gg 1$ has nothing to do with random or chaotic properties of the system.
It means only that due to well known phenomenom of level repulsion the
two level approximation is not adequate. Lorentzian envelope (see Fig. 3) we found
arises from the interaction of a single level in $L$-well with a set of levels
in the $R$-well and not with appearance of level widths (imaginary self-energy
contributions).

One should distinguish between short-time and long-time behavior, and
the boundary between them depends on the parameter $\Lambda $.
Short-time returns ($\propto \beta $) are governed by one or a small number of semiclassical
paths, while long-time returns ($\propto R_n^{-1}$) arise from interference between many paths.
In the limit $\Lambda \ll 1$, exponential decay occurs for short-time dynamics,
while the system remains regular for long-time scales, in contrast with chaotic
models we discussed in the Introduction. Nevertheless the tunneling in the limit
of $\Lambda \gg 1$ can induce vibrational relaxation for localized
$R$-levels. The relaxation appears due to tunneling recurrences, and results
in redistribution of initial energy over all levels coupled with a single $L$-level.

Main physical idea of our paper, namely that specific quasi-chaotic behavior
is associated with the fact that one level in $L$- well in a certain condition
($\Lambda \gg 1$) is coupled to a set of almost dense levels in the $R$-well,
was discussed in the literature long ago \cite{HO55} (see also
\cite{zeld}), mainly qualitatively.
Our achievement is that we alone seem to have propose the concrete and 
tractable analytical
model to illustrate and to investigate explicitely and quantitatively this 
statement.

In this respect our results are quite different from numerical investigations
of billiard-type systems (see e.g. review article \cite{BT93}), showing
universal behavior of level spacings in finite chaotic systems.
Our results (for the totally integrable 1d model) demonstrate
that level spacing distribution is not a specific feature of quantum
systems with chaotic classical counterpart limit. Our finding
of the equidistant regular level distribution is a result
of the interaction of the single $L$-level with several
(of the order of 10 for our particular choice of the parameters)
$R$-levels (which in own turn are regular ones).
As well we should distinguish our model and dynamic tunneling
ones \cite{WI88}, \cite{FD98}. The latter assumed strong coupling of the
tunneling system with an environment which destroys the coherence,
whereas in our model the coherence is destroyed by the tunneling itself due
to the high density of $R$-states, breaking two level approximation.

Note also at the very end of the paper that results presented here
not only interested in their own right (at least in our opinion)
but they might be directly tested experimentally since there
are many molecular systems where investigated in the paper 1d 
asymmetric potential
is a reasonable model for the reality. And not only molecular systems, 
for instance recently as a controllable two-level system, double
quantum dots are proposed for realizing 
a single quantum bit in solid state systems. Experimentally
\cite{YF99} in these
systems there observed two distinct regimes characterizing the nature
of low-energy dynamics:

(i) relaxational regime, when an excited-state electron population decays
exponentially in time with a rate correctly given by Fermi golden rule;

(ii) vibronic regime, when the population oscillates for some number
of cycles before decaying.

And what's more, at short times the averaged excited-state populations
oscillates but has a decaying envelope. The similarity with the
behavior we found in the paper is evident.\footnote{All characteristics
of our model are not specific only for 1d case. For $\Lambda \gg 1$
one can expect similar behavior and for multidimensional systems.}

\acknowledgements
The research described in
this publication was made possible in part by RFFR Grants 97-03-33687a and 00-02-11785.
The numerical results origin from a collaboration with E.V.Vetoshkin
whose contribution is greatly acknowledged. Also we express our sincere
gratitude to the referee who read our manuscript and made a number
of valuable comments.

\appendix \section*{}
The semi-classical wave function is represented in the well known WKB form \footnote{
Equivalently it can be represented in the so-called instanton or minimum
action tunneling path formalism \cite{pol} (see also \cite{b1}) in the form of $\Psi = 
\exp (-\gamma W_E)$, which is more efficient for classically inaccessible parts of phase space.}
$$
\Psi = \exp \left (i W\right ) \, ,
$$
The action $W$ should satisfy to WKB equation
\begin{eqnarray}
\label{a1}
\frac{1}{2} \left (\frac{d W}{d X}\right )^2 = \frac{\varepsilon }{\gamma } - V(X) \, ,
\end{eqnarray}
and two turning points, which are boundaries of classically allowed regions, are
situated near zeros of $V(X) - \varepsilon /\gamma $.

For the asymmetric double-well potential (\ref{b1}) the Bohr - Sommerfeld
\cite{LL} quantization equations read
\begin{eqnarray}
\label{A1}
tg (\gamma W_L)tg (\gamma W_R) = 4 \exp(2\gamma W_B) \, ,
\end{eqnarray}
where $W_B$ is the action in the classically forbidden region in between the
turning points $X_1 , X_2$ in the left and right wells, and $W_{L , R}$ are
the coordinate independent actions in the classically allowed regions
inside of the $L$ (respectively $R$) well.
Using the following expansion
$$
\tan z = \sum _{m=0}^{\infty } 2 z\left [z^2 - \pi ^2\left (m +\frac{1}{2}^2\right )\right ]^{-1} \, ,
$$
one gets the almost equidistant spectrum of the mixed $L-R$ levels, and in this condition
the solution of (\ref{A1}) leads to the expressions (\ref{b9}), (\ref{b10}) 
presented in the main text of the paper. 

The time evolution of any initially prepared state can be described by a superposition
of the eigenfunctions of the discrete and continuous spectra with time dependent phases.
For the potential (\ref{b1}) with $b \gg 1$ the initial finite motion,
i.e.
the initial density distribution
\begin{eqnarray}
\label{n1}
\rho (t) = \int _{X_1}^{X_2} |\Psi (X , t)|^2 d X  \,
\end{eqnarray}
concentrated in the $L$-well at $t=0$ decreases exponentially with time
\begin{eqnarray}
\label{n2}
\rho (t) = \rho (0) \exp \left (-\eta t\right )  \, .
\end{eqnarray}
Eq. (\ref{n2}) signifies that the wave functions of quasi-stationary
states have the form
\begin{eqnarray}
\label{n3}
\Psi _n(X ,t) = \Psi _n(X)  \exp \left (\left (-i \varepsilon _n -  \eta _n/2 \right )t\right )  \, ,
\end{eqnarray}
and the eigenvalues are complex and lies on the lower half-space of $(\varepsilon , \eta )$ plane. 
The quantization of the stationary states of a discrete spectrum is performed by the requirement
\cite{LL}
$$
|\Psi (X , t)|^2 \to 0 \, , \, {\rm at} \, |X| \to \infty \, .
$$
This condition is impossible to impose 
to quasi-stationary states, since the wave
functions $\Psi _n(X)$ exponentially is increased in the region of $\varepsilon \gg V(X)$.
The physically meaningful boundary condition as was noted first by Zeldovich \cite{zeld}
for quasi-stationary states can be written as a conservation law
for the flux probability from the $L$-well through the barrier. 
The difference between stationary and quasi-stationary states disappears as
it should be at $\eta \to 0$.

The expansion of the initially quasi-stationary state is dominated by
the continuum spectrum eigenfunctions with the energies close
to the real parts of the eigenvalues $\varepsilon _n$.
These eigenfunctions have the form
\begin{eqnarray}
\label{n4}
\Psi _k(X) = \left ( 
\begin{array}{cc}
A(k)\phi _k^0 (X) \, , \quad   X < X_m  \\ 
\sqrt {\frac{2}{\pi }} \sin(k X + \delta (k)) \, , \quad X > X_m 
\end{array} 
\right ) \, ,
\end{eqnarray}
where $X_m$ is the left turning point of the $R$-well, the localized wave
function $\phi _k^0$
is normalized to unity, and the phase is given
\begin{eqnarray}
\label{n5}
\delta (k) = \delta _0 - \arctan \frac{k_2}{k-k_1}  \, ,
\end{eqnarray}
and $\delta _0$ is $k$-independent component, $k_1 = \sqrt {2 m \varepsilon _n} $,
$k_2 = k_1\eta _n/4\varepsilon _n$.
For the eigenfunctions with the energies $\varepsilon $ and $\varepsilon ^\prime $ close to
$\varepsilon _n$ we get
\begin{eqnarray}
\label{n6}
\int _{-\infty }^{X} \phi _k(X^\prime )\phi _{k^\prime }(X^\prime ) d X^\prime =
\frac{1}{2 m}\left (\frac{1}{\varepsilon - \varepsilon ^\prime }\right )\left (
\phi _k^\prime \frac{d\phi _k}{dX} - \phi _k^\prime \frac{d \phi _k^\prime }{d X}\right )
\, .
\end{eqnarray}
From (\ref{n4}), (\ref{n5}), and (\ref{n6}) in the limit $\varepsilon - \varepsilon ^\prime \to 0$ we get
\begin{eqnarray}
\label{n7}
A^2(k) = \frac{2}{\pi } \sqrt {\frac{2 \varepsilon _n}{m}} \frac{\eta _n}{4(\varepsilon - \varepsilon
_n)^2 + \eta _n^2} \, .
\end{eqnarray}
Expressions (\ref{n5}), (\ref{n7}) are valid for a continuous spectrum,
for discrete levels the phase shift as well is governed by the probability
flux from the $R$-well into classically forbidden region, and instead of
(\ref{n5}) it leads
\begin{eqnarray}
\label{n9}
\delta  = \arctan \sqrt{{R_n}{\beta }}\frac{1}{\varepsilon _n - \varepsilon
_{nm}} \, ,
\end{eqnarray}
and instead of (\ref{n7}) one can easily find
\begin{eqnarray}
\label{n10}
A^2(\varepsilon _{nm}) = \frac{2}{\pi }\frac{\sqrt 
{R_n}}{\beta (\varepsilon _n - \varepsilon _{nm})^2 + R_n}
\, , 
\end{eqnarray}
Note that (\ref{n10}) has almost the same form as (\ref{n7}), although it
depends on discrete energy levels, and besides it has a different
coefficient due to different normalization condition.

The relation (\ref{n7}) shows that the probability density
of the continuous spectrum
eigenstates exhibits the Lorentzian distribution around the real part 
of the quasi-stationary
eigenvalues $\varepsilon _n$. Expressions (\ref{n7}) -(\ref{n10}) are
equivalent to the spectral distribution (\ref{b18}) presented in the main body of the paper.

Few words concerning numerical results presented in the main text in 
the figures 1 - 3.
The calculations have been performed to check:

(i) semiclassical approximation for the model potential (\ref{b1});

(ii)
the spectral distribution (\ref{b18}).

We used the numerical diagonalization of the Hamiltonian matrix in the basis set
of trial functions, which includes: so-called instanton wave functions of the
$L$-well (see \cite{b1}), and the WKB functions of $R$-well. This basis
was orthonormalized by using standard Schmidt method \cite{morse}.
For the $L$-well highly excited states near the barrier top have been
also included. In all numerical calculations we set the value of $\alpha _0$ 
(so-called defect of a resonance) as zero. All results presented on the figures
do not depend on this particular choice.

The numerical results confirm that Eq. (\ref{b18}) is quiet
accurate in the whole range of $\Lambda $ where the transition from coherent
oscillations to exponential decay occurs.
Note that since $R$-levels with the negative energy are not mixed with $L$-levels,
and besides the resonance region is sufficiently narrow ($R_n =0.01$), we need not
diagonalize huge matrices. For our purposes the diagonalization of the
matrix $3000 \times 3000$ is more than sufficient to find eigenvalues in the resonance
region around the $n=0$ $L$-level.

\newpage

\centerline{Figure Captions.}

Fig. 1

The eigenvalues 
as functions of $\Lambda $ for the zero-point level ($n=0$) of 
the $L$-well. Dashed lines indicate the limits of $\Lambda  \ll 1 $, and $\Lambda \gg
1$;
$\gamma = 10$ , $\alpha _0 =0$.

Fig. 2

The survival probability for different
values of $\Lambda $ and $\gamma = 10$ : 

(a) $\Lambda  = 0.02$ ,  $b=5$ (solid line) ; $\Lambda  = 0.5$ ,  $b=116$ (dashed line)  ;

(b) $\Lambda = 0.5$ ,  $b=116$ (solid line) ;  $\Lambda  = 4.0$ , $b=929$  (dashed line) ; 

(c) $\Lambda  = 4.0$ , $b=929$ (solid line) ; $\Lambda  = 16.0$ , $b=3715$ (dashed line) .

Fig. 3

The spectral distribution for
different values of 
$\Lambda $  
and $\gamma = 10 $: 

(a) $\Lambda  = 0.02$ , $b=5$ ; 

(b) $\Lambda  = 4.0$ , $b=929$ ;

(c) $\Lambda  = 20.0$ , $b=4644$ .

\newpage


\begin{references} 
\bibitem{YF99} 
H.S.Yang, S.P.Feofilov, D.K.Williams, 
et al., 
Physica B, {\bf 263}, 476 (1999).
\bibitem{ZC79} G.M.Zaslavsky, Phys. Repts., {\bf 80}, 157 (1980). 
\bibitem{LL83} A.J.Lichtenberg,
M.A.Lieberman, Regular and Stochastic Motion, Springer, New York (1983).
\bibitem{WJ82} Y.Weissman, J.Jortner, J. Chem. Phys., {\bf 77}, 1469 (1982);
ibid., 1486.
\bibitem{BE1} M.V.Berry, in Chaotic Behavior of Deterministic Systems, ed. by
G.Iooss, R.Helleman, R.Stora, North-Holland, Amsterdam (1983).
\bibitem{BE2} M.V.Berry, Proc. Royal .Soc., A, {\bf 423}, 219 (1989).
\bibitem{NA} K.Nakamura, Quantum Chaos, Kluver, Dordrecht (1997).
\bibitem{KH00} L.Kaplan, E.J.Heller, Phys. Rev. E, {\bf 62}, 409 (2000).
\bibitem{FPU} E.Fermi, J.Pasta, S.Ulam, Los Alamos Scientific Laboratory
Report, LA-1940 (1955). 
\bibitem{JA63} E.A.Jackson, J. of Math. Phys.,
{\bf 4}, 686 (1963). 
\bibitem{FW63} J.Ford, J.Waters, J. of Math. Phys.,
{\bf 4}, 1293 (1963). 
\bibitem{BR84} M.V.Berry, M.Robnik, J. Phys. A, {\bf 17}, 2413 (1984).
\bibitem{CG86} E.Caurier, B.Grammaticos, Europhys. Lett., {\bf 2}, 417 (1986).
\bibitem{ZM86} T.Zimmermann, H.D.Meyer, H.Koppel, L.S.Cederbaum, Phys. Rev. A,
{\bf 33}, 4334 (1986).
\bibitem{TU94} S.Tomsovic, D.Ullmo, Phys. Rev. E., {\bf 50}, 145 (1994).
\bibitem{SW01} D.A.Steck, W.H.Windell, M.G.Raizen, Science, {\bf 293}, 274 (2001).
\bibitem{BT93} O.Bohigas, S.Tomsovich, D.Ullmo, Phys. Repts., 
{\bf 223}, 43 (1993).
\bibitem{BV99} V.A.Benderskii, E.V.Vetoshkin, H.P.Trommsdorf, Chem. Phys., {\bf 244},
273 (1999).
\bibitem{LL} L.D.Landau, E.M.Lifshits, Quantum Mechanics (non-relativistic
theory), Pergamon Press, New York (1965). 
\bibitem{b1} V.A.Benderskii, E.V.Vetoshkin, 
Chem. Phys., {\bf 257}, 203 (2000). 
\bibitem{zeld} Ya.B.Zeldovich, JETP, {\bf 12}, 542 (1961).
\bibitem{HO55} L.van Hove, Physica, {\bf 21}, 517 (1955).
\bibitem{WI88} M.Wilkinson, J. Phys. A., {\bf 21}, 1173 (1988).
\bibitem{FD98} S.D.Frischat, E.Doron, Phys. Rev. E., {\bf 57}, 1421 (1998).
\bibitem{pol} A.M.Polyakov, Nucl.Phys. B, {\bf 129},
429 (1977). 
\bibitem{morse} P.M.Morse, M.Feshbach, Methods of Theoretical Physics, 
Mc.Graw Hill, N.Y. (1953).
\end{references}
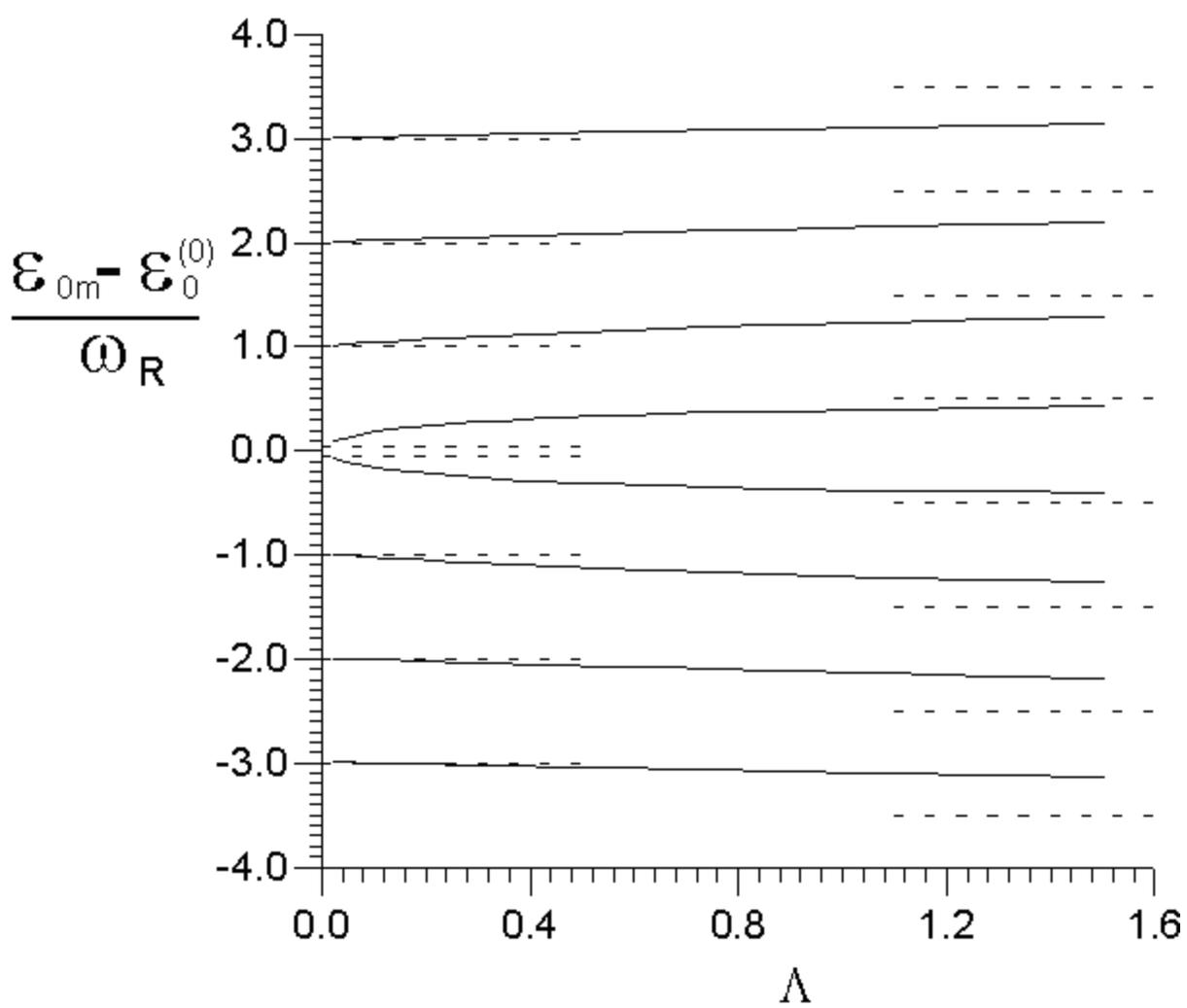
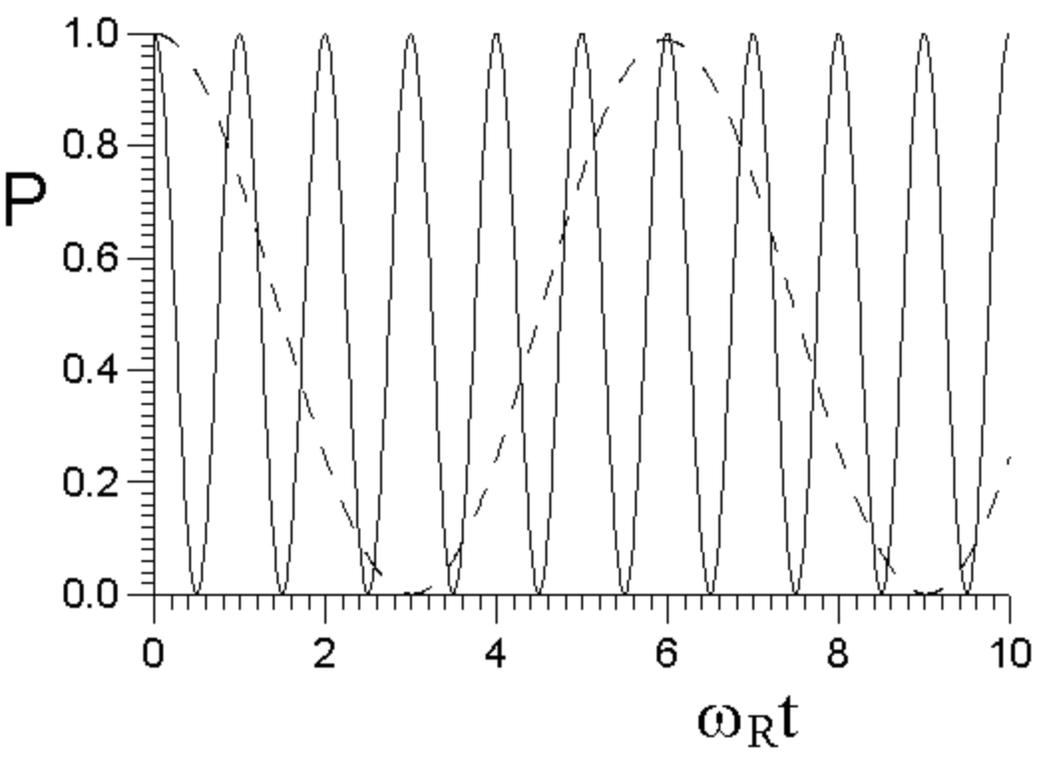
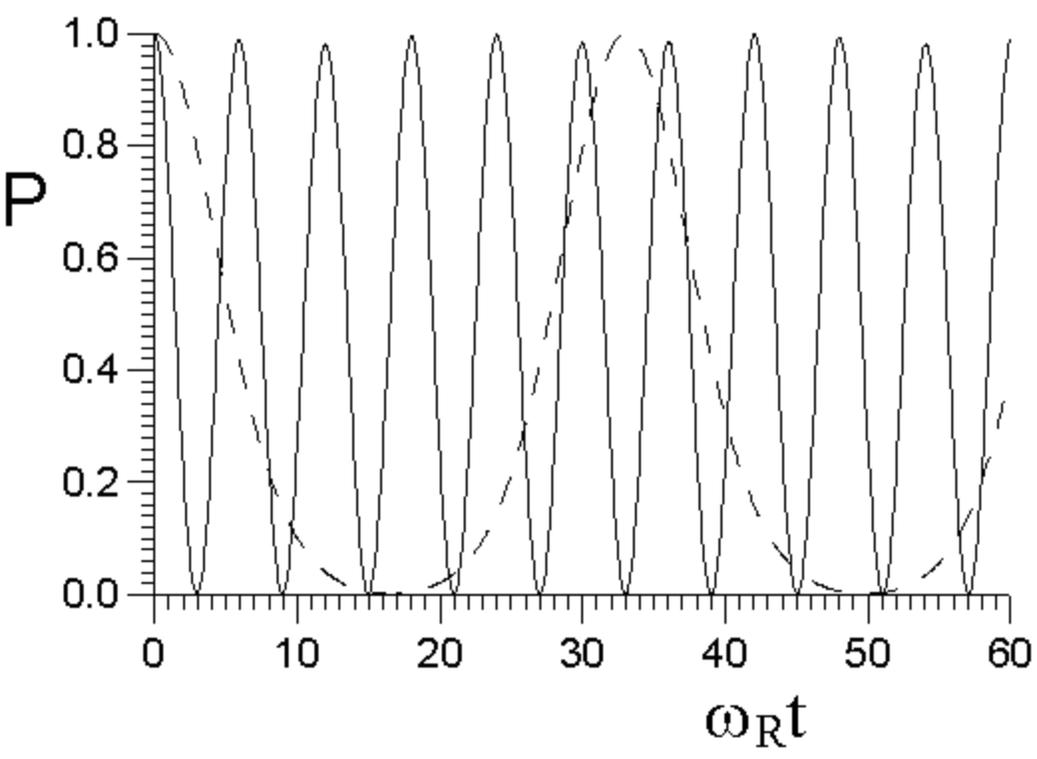
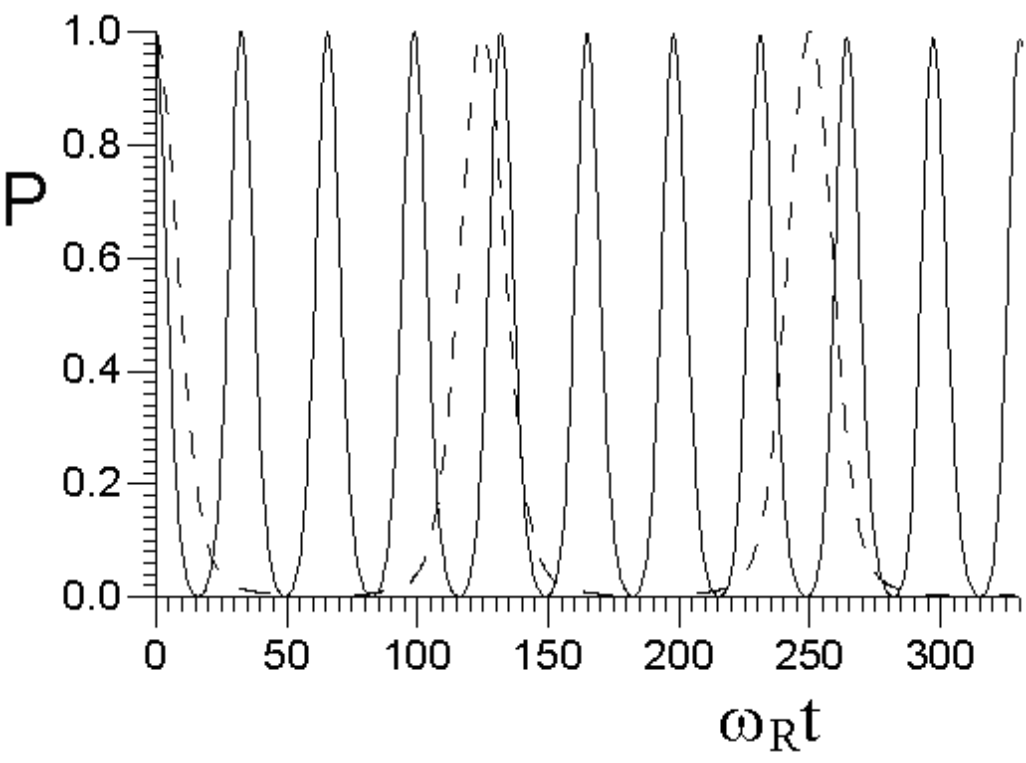
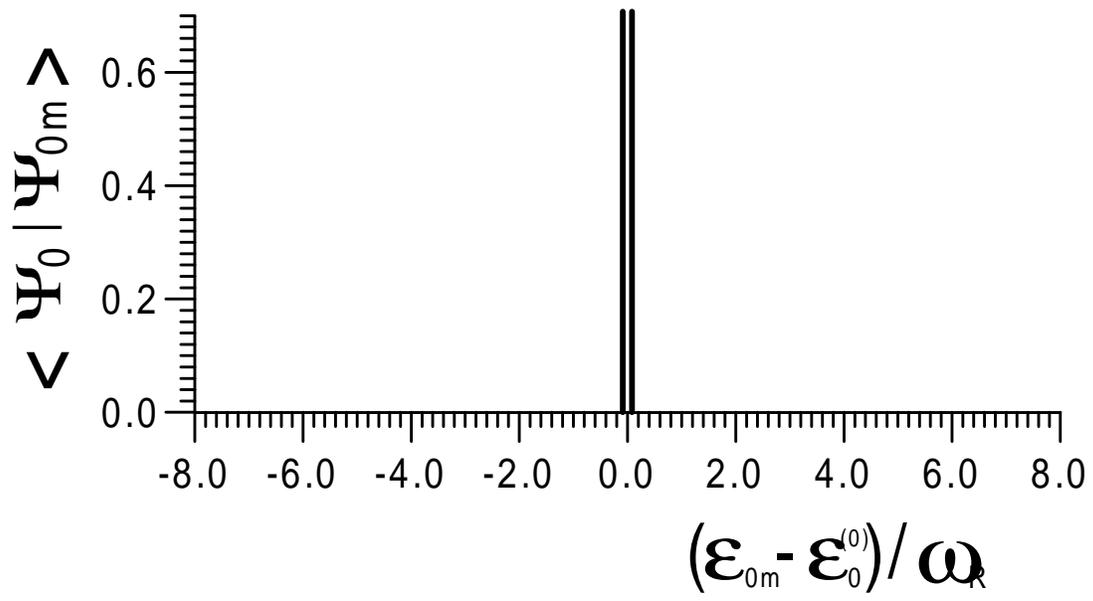
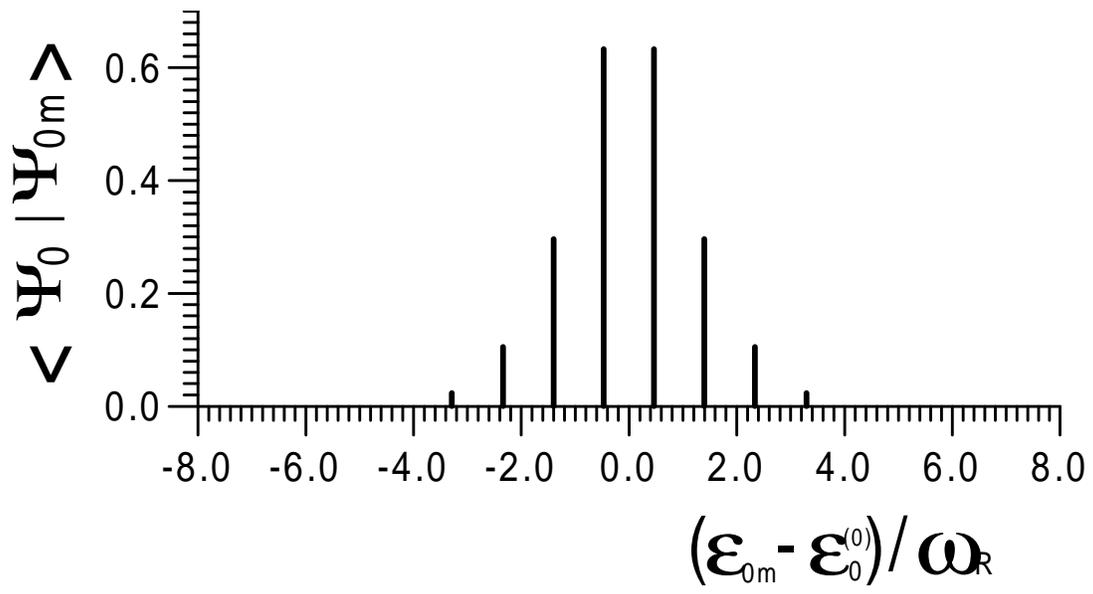
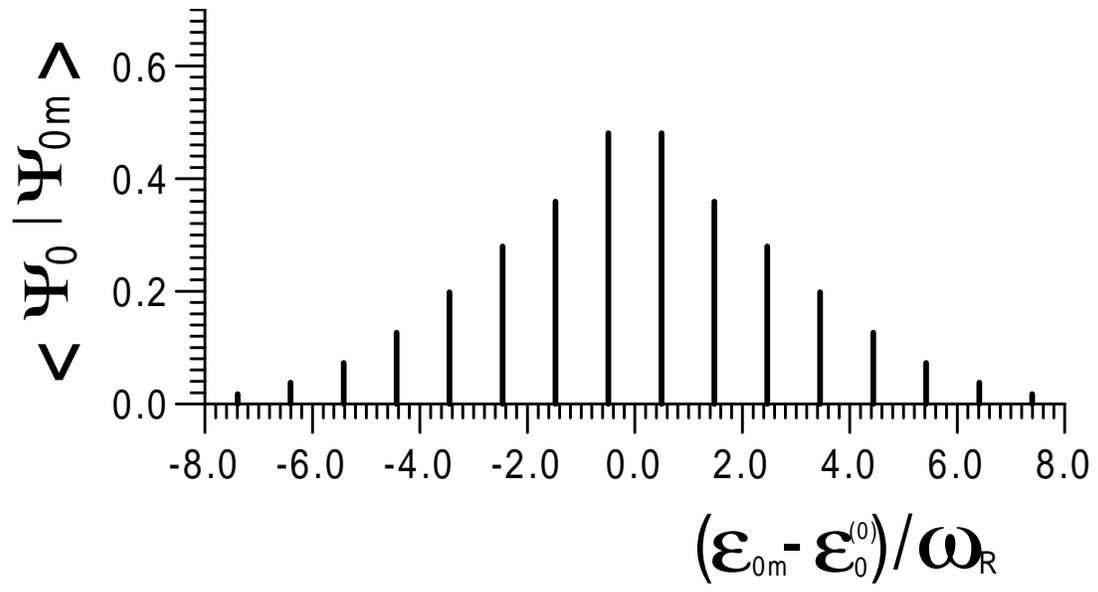
\end{document}